# Redistribution Mechanisms for Assignment of Heterogeneous Objects

**Sujit Gujar**                                        SUJIT@CSA.IISC.ERNET.IN
**Y Narahari**                                         HARI@CSA.IISC.ERNET.IN
*Dept of Computer Science and Automation*
*Indian Institute of Science, Bangalore, 560012*

## Abstract

There are $p$ heterogeneous objects to be assigned to $n$ competing agents $(n > p)$ each with unit demand. It is required to design a Groves mechanism for this assignment problem satisfying weak budget balance, individual rationality, and minimizing the budget imbalance. This calls for designing an appropriate rebate function. When the objects are identical, this problem has been solved which we refer as WCO mechanism. We measure the performance of such mechanisms by the redistribution index. We first prove an impossibility theorem which rules out linear rebate functions with non-zero redistribution index in heterogeneous object assignment. Motivated by this theorem, we explore two approaches to get around this impossibility. In the first approach, we show that linear rebate functions with non-zero redistribution index are possible when the valuations for the objects have a certain type of relationship and we design a mechanism with linear rebate function that is worst case optimal. In the second approach, we show that rebate functions with non-zero efficiency are possible if linearity is relaxed. We extend the rebate functions of the WCO mechanism to heterogeneous objects assignment and conjecture them to be worst case optimal.

## 1. Introduction

Consider that $p$ resources are available and each of $n > p$ agents is interested in utilizing one of them. It is desirable that we assign the resources to the agents who value them the most. Since the classical Vickery-Clarke-Groves mechanisms (Vickrey, 1961; Clarke, 1971; Groves, 1973) have attractive properties such as dominant strategy incentive compatibility (DSIC) and allocative efficiency (AE), Groves mechanisms are quite appealing to use in this context. However, in general, a Groves mechanism need not be budget balanced. That is, the total transfer of money in the system may not be zero. So the system will be left with a surplus or deficit. Using Clarke's (1971) mechanism, we can ensure under fairly weak conditions, that there is no deficit of money (that is the mechanism is weakly budget balanced). In such a case, the system or the auctioneer will be left with some money.

Often, the surplus money is not really needed in many social settings such as allocations by the Government among its departments, etc. Since strict budget balance cannot coexist with DSIC and AE (Green-Laffont theorem, see Green & Laffont, 1979), we would like to redistribute the surplus to the participants as far as possible, preserving DSIC and AE. This idea was originally proposed by Laffont (1979). The total payment made by the mechanism as a redistribution will be referred to as the *rebate* to the agents.

In this paper, we consider the following problem. There are $n$ agents and $p$ heterogeneous objects ($n > p > 1$). Each agent desires one object out of these $p$ objects. Each agent's valuation for any of the objects is independent of his valuations for the other objects. Valuations of the different agents are also mutually independent. Our goal is to design a mechanism for assignment of the $p$ objects among the $n$ agents which is allocatively efficient, dominant strategy incentive compatible, and maximizes the rebate (which is equivalent to minimizing the budget imbalance). In addition, we would like the mechanism to satisfy feasibility and individual rationality. Thus, we seek to design a Groves mechanism for assigning $p$ heterogeneous objects among $n$ agents satisfying:

1. Feasibility (F) or weak budget balance. That is, the total payment to the agents should be less than or equal to the total received payment.

2. Individual Rationality (IR), which means that each agent's utility by participating in the mechanism should be non-negative.

3. Minimizes budget imbalance.

We call such a Groves mechanism that redistributes Clarke's Payment as *Groves redistribution mechanism* or simply *redistribution mechanism*. Designing a redistribution mechanism involves the design of an appropriate *rebate function*. If in a redistribution mechanism, the rebate function for each agent is a linear function of the valuations of the remaining agents, we refer to such a mechanism as a *linear redistribution mechanism* (LRM). In many situations, design of an appropriate LRM reduces to a problem of solving a linear program.

Due to the Green-Laffont theorem , we cannot guarantee 100% redistribution for all type profiles. So a performance index for the redistribution mechanism would be the worst case redistribution, that is, the fraction of the surplus which is guaranteed to be redistributed irrespective of the bid profiles. This fraction will be referred to as *redistribution index* in the rest of the paper. The advantage of worst case analysis is that, it does not require any distributional information on the type sets of the agents. It is desirable that the rebate function is deterministic and anonymous. A rebate function is said to be anonymous if two agents having the same bids get the same rebate. Also, when valuation spaces are identical for all the agents, without loss of generality, we can restrict our attention to the anonymous rebate functions. Thus, the aim is to design an anonymous, deterministic rebate function which maximizes the redistribution index and satisfies feasibility and individual rationality.

Our work in this paper seeks to non-trivially extend the results of Moulin (2009) and Guo and Conitzer (2009) who have independently designed a Groves mechanism in order to redistribute the surplus when objects are identical (homogeneous objects case). Their mechanism is deterministic, anonymous, and has maximum redistribution index over all possible Groves redistribution mechanisms. We will refer to their mechanism as the *worst case optimal* (WCO) mechanism. The WCO Mechanism is a linear redistribution mechanism. In this paper, we concentrate on designing a linear redistribution mechanism for the heterogeneous objects case.

## 1.1 Relevant Work

As it is impossible to achieve allocative efficiency, DSIC, and strict budget balance simultaneously, we have to compromise on one of these properties. Faltings (2005) and Guo and Conitzer (2008a) achieve budget balance by compromising on AE. If we are interested in preserving AE and DSIC, we have to settle for a non-zero surplus or a non-zero deficit of the money (budget imbalance) in the system. To reduce budget imbalance, various rebate functions have been designed by Bailey (1997), Cavallo (2006), Moulin (2009), and Guo and Conitzer (2009). Moulin (2009) and Guo and Conitzer (2009) designed a Groves redistribution mechanism for assignment of $p$ homogeneous objects among $n > p$ agents with unit demand. Guo and Conitzer (2009) generalize the work in earlier paper (Guo & Conitzer, 2007) for multi-unit demand of identical items. In the work of Guo and Conitzer (2008b), authors designed a redistribution mechanism which is optimal in the expected sense for the homogeneous objects setting. Thus, it will require some distributional information over the type sets of the agents. Clippel and co-authors (2009) use the idea of destroying some of the items to maximize the agents' utilities. A preliminary version of the results presented in this paper have appeared in our earlier papers (Gujar & Narahari, 2009, 2008).

## 1.2 Contributions and Outline

Our objective in this paper is to design a Groves redistribution mechanism for assignment of heterogeneous objects with unit demand. To the best of our knowledge, this is the first attempt to design a redistribution mechanism for assignment of heterogeneous objects.

First, we investigate the question of existence of a linear rebate function for redistribution of surplus in assignment of heterogeneous objects. Our result shows that in general, when the domain of valuations for each agent is $\mathbb{R}_+^p$, it is impossible to design a linear rebate function, with non-zero redistribution index, for the heterogeneous settings. However, we can relax the assumption of independence of valuations of different objects to get a linear rebate function with non-zero redistribution index. Another way to get around the impossibility theorem is to relax the linearity requirement of a rebate function. In particular, our contributions in this paper can be summarized as follows.

- We first prove the impossibility of existence of a linear rebate function with non-zero redistribution index for heterogeneous settings, when the domain of valuations for each agent is $\mathbb{R}_+^p$ and the valuations for the objects are independent.

- When the objects are heterogeneous but the values for the objects of an agent can be derived from one single number, we design a Groves redistribution mechanism that is linear, anonymous, deterministic, feasible, individually rational, and efficient. In addition, the mechanism is worst case optimal with non-zero redistribution index.

- We show the existence of a non-linear rebate function that has a non-zero redistribution index.

- We propose a mechanism, HETERO, which extends Moulin/WCO mechanism for heterogeneous settings. We conjecture HETERO to have non-zero redistribution index and to be worst case optimal.

The paper is organized as follows. In Section 2, we introduce the notation followed in the paper and describe relevant background work from the literature. We also explain the WCO mechanism there. In Section 3, we state and prove our impossibility result. We derive an extension of the WCO mechanism for heterogeneous objects but with single dimensional private information in Section 4. The impossibility result does not rule out possibility of non-linear rebate functions with strictly positive redistribution index. We show this with a redistribution mechanism, BAILEY-CAVALLO, which is Bailey's mechanism (1997) applied to the settings under consideration in Section 5. We design another non-linear rebate function, HETERO, that actually matches with Moulin's rebate function when the objects are identical. We describe the construction of HETERO in Section 5. We have carried out simulations to provide empirical evidence for our conjecture regarding HETERO. The experimental setup and results are described in Section 6. We conclude the paper in Section 7 and provide some directions for future work. In our analysis, we need an ordering of the bids of the agents which we define in Appendix A. The proofs of some of the lemmas in the paper are presented in Appendix B.

## 2. Preliminaries and Notation

In this section we will first define the notation used in the paper and preliminaries about the redistribution mechanisms.

### 2.1 The Model and Notation

The notation used is summarized in Table 1. Where the context is clear, we will use $t, t_i, r_i, k,$ and $v_i$ to indicate $t(b), t_i(b), r_i(b), k(b),$ and $v_i(k(b))$ respectively. In this paper, we assume that the payment made by agent $i$ is of the form $t_i(\cdot) - r_i(\cdot)$, where $t_i(\cdot)$ is agent $i$'s payment in the Clarke pivotal mechanism (1971). We refer to $\sum_i t_i$, as the total Clarke payment or the surplus in the system.

In general, we assume there are $n$ agents and $p$ distinct objects. We also assume that the allocation rule satisfies the allocative efficiency (AE) property.

### 2.2 Important Definitions

We provide a few important definitions here in a conceptual way.

**Definition 1 (DSIC)** *We say a mechanism is* Dominant Strategy Incentive Compatible *(DSIC) if it is a best response for each agent to report its type truthfully, irrespective of the types reported by the other agents.*

**Definition 2 (Allocative Efficiency)** *We say a mechanism is* allocatively efficient *(AE) if the mechanism chooses, in every given type profile, an allocation of objects among the agents such that sum of the valuations[1] of the allocated agents is maximized.*

**Definition 3 (Redistribution Mechanism)** *We refer to a Groves mechanism as a* Groves redistribution mechanism *or simply* redistribution mechanism, *if it allocates objects to the* agents in an allocatively efficient way and redistributes the Clarke surplus in the system in the form of rebates to the agents such that the net payment made by each agent still follows the Groves payment structure.

---

1. Sum of the valuations of the allocated agents in an allocation is also referred as total value or value of the allocation.

| | |
|---|---|
| $n$ | Number of agents |
| $N$ | Set of the agents $= \{1, 2, \ldots, n\}$ |
| $p$ | Number of objects |
| $i$ | Index for an agent, $i = 1, 2, \ldots, n$ |
| $j$ | Index for object, $j = 1, 2, \ldots, p$ |
| $\mathbb{R}_+$ | Set of positive real numbers |
| $\Theta_i$ | The space of valuations of agent $i$, $\Theta_i = \mathbb{R}_+^p$ |
| $b_i$ | Bid submitted by agent $i$, $= (b_{i1}, b_{i2}, \ldots, b_{ip}) \in \Theta_i$ |
| $b$ | $(b_1, b_2, \ldots, b_n)$, the bid vector |
| $K$ | The set of all allocations of $p$ objects to $n$ agents, each getting at most one object |
| $k(b)$ | An allocation, $k(\cdot) \in K$, corresponding to the bid profile b |
| $k^*(b)$ | An allocatively efficient allocation when the bid profile is $b$ |
| $k^*_{-i}(b)$ | An allocatively efficient allocation when the bid profile is $b$ and agent $i$ is excluded from the system |
| $v_i(k(b))$ | Valuation of the allocation $k$ to the agent $i$, when $b$ is the bid profile |
| $v$ | $v : K \to \mathbb{R}$, the valuation function, $v(k(b)) = \sum_{i \in N} v_i(k(b))$ |
| $t_i(b)$ | Payment made by agent $i$ in the Clarke pivotal mechanism, when the bid profile is b, $t_i(b) = v_i(k^*(b)) - \left(v(k^*(b)) - v(k^*_{-i}(b))\right)$ |
| $t(b)$ | The Clarke payment, that is, the total payment received from all the agents, $t(b) = \sum_{i \in N} t_i(b)$ |
| $t^{-i}$ | The Clarke payment received in the absence of the agent $i$ |
| $r_i(b)$ | Rebate to agent $i$ when bid profile is $b$ |
| $e$ | The redistribution index of the mechanism, $= \inf_{b:t(b) \neq 0} \frac{\sum r_i(b)}{t(b)}$ |

Table 1: Notation: redistribution mechanisms

**Definition 4 (Linear Rebate Function)** *We say the rebates to an agent follow a linear rebate function if the rebate is a linear combination of bid vectors of all the remaining agents. Moreover, if a redistribution mechanism uses linear rebate functions for all the agents, we say the mechanism is a linear redistribution mechanism.*

**Definition 5 (Redistribution Index)** *The redistribution index of a redistribution mechanism is defined to be the worst case fraction of Clarke's surplus that gets redistributed among the agents. That is,*

$$e = \inf_{b:t(b) \neq 0} \frac{\sum r_i(b)}{t(b)}$$

## 2.3 Optimal Worst Case Redistribution when Objects are Identical

When the objects are identical, every agent $i$ has the same value for each object, call it $v_i$. Without loss of generality, we will assume, $v_1 \geq v_2 \geq \ldots \geq v_n$. In Clarke's pivotal mechanism, the first $p$ agents will receive the objects and each of these $p$ agents will pay $v_{p+1}$. So, the surplus in the system is $pv_{p+1}$. For this situation, Moulin (2009) and Guo and Conitzer (2009) have independently designed a redistribution mechanism.

Guo and Conitzer (2009) maximize the worst case fraction of the total surplus which gets redistributed. This mechanism is called the WCO mechanism. Moulin (2009) minimizes the ratio of budget imbalance to the value of an optimal allocation, that is the value of an allocatively efficient allocation. The WCO mechanism coincides with Moulin's feasible and individually rational mechanism. Both the above mechanisms work as follows. After receiving bids from the agents, bids are sorted in decreasing order. The first $p$ agents receive the objects. Each agent's Clarke payment is calculated, say $t_i$. Every agent $i$ pays, $p_i = t_i - r_i$, where, $r_i$ is the rebate function for an agent $i$.

$$
\begin{array}{lll}
r_i^{WCO} = & c_{p+1}v_{p+2} + c_{p+2}v_{p+3} + \ldots + c_{n-1}v_n & i = 1, \ldots p+1 \\
r_i^{WCO} = & c_{p+1}v_{p+1} + \ldots + c_{i-1}v_{i-1} + c_i v_{i+1} + \ldots + c_{n-1}v_n & i = p+2, \ldots n
\end{array} \tag{1}
$$

where,

$$
c_i = \frac{(-1)^{i+p-1}(n-p)\binom{n-1}{p-1}}{i\binom{n-1}{i}\sum_{j=p}^{n-1}\binom{n-1}{j}}\left\{\sum_{j=i}^{n-1}\binom{n-1}{j}\right\}; \quad i = p+1, \ldots, n-1 \tag{2}
$$

Suppose $y_1 \geq y_2 \geq \ldots \geq y_{n-1}$ are the bids of the $(n-1)$ agents excluding the agent $i$, then equivalently the rebate to the agent $i$ is given by,

$$
r_i^{WCO} = \sum_{j=p+1}^{n-1} c_j y_j \tag{3}
$$

The redistribution index of this mechanism is $e^*$, where $e^*$ is given by,

$$
e^* = 1 - \frac{\binom{n-1}{p}}{\sum_{j=p}^{n-1}\binom{n-1}{j}}
$$

This is an optimal mechanism, since there is no other mechanism which can guarantee more than $e^*$ fraction redistribution in the worst case.

Before we proceed to present our impossibility theorem we state the following theorem by Guo and Conitzer (2009) which will be used to design our mechanism.

**Theorem 1** *(Guo & Conitzer, 2009) For any $x_1 \geq x_2 \geq \ldots x_n \geq 0$,*

$$
a_1 x_1 + a_2 x_2 + \ldots a_n x_n \geq 0 \ \ iff \ \ \sum_{i=1}^{j} a_i \geq 0 \ \ \forall j = 1, 2 \ldots, n
$$

## 3. Impossibility of Linear Rebate Function with Non-Zero Redistribution Index

We have just reviewed the design of a redistribution mechanism for homogeneous objects. We have seen that the WCO mechanism is a linear function of the types of agents. We now explore the general case. In the homogeneous case, the bids are real numbers which can be arranged in decreasing order. The Clarke surplus is a linear function of these ordered bids. For the heterogeneous scenario, this would not be the case. Each bid $b_i$ belongs to $\mathbb{R}_+^p$; hence, there is no unique way of defining an order among the bids. Moreover, the Clarke surplus is not a linear function of the received bids in the heterogeneous case. So, we cannot expect any linear/affine rebate function of types to work well at all type profiles. We will prove this formally.

We first generalize a theorem from the work of Guo and Conitzer (2009). The context in which Guo and Conitzer stated and proved the theorem is in the homogeneous setting. We show that this result holds true in the heterogeneous objects case also. The symbol $\succcurlyeq$ denotes the order over the bids of the agents, as defined in the A.2.

**Theorem 2** *In the Groves redistribution mechanism, any deterministic, anonymous rebate function $f$ is DSIC iff,*

$$r_i = f(v_1, v_2, \ldots, v_{i-1}, v_{i+1}, \ldots, v_n) \ \ \forall i \in N \tag{4}$$

*where, $v_1 \succcurlyeq v_2 \succcurlyeq \ldots \succcurlyeq v_n$.*

**Proof:**

- The "if" part: If $r_i$ takes the form given by equation (4), then the rebate of agent $i$ is independent of his valuation. The allocation rule satisfies allocative efficiency. So, the mechanism is still Groves and hence DSIC. The rebate function defined is deterministic. If two agents have the same bids, then, as per the ordering defined in Appendix, $\succcurlyeq$, they will have the same ranking. Suppose agents $i$ and $i+1$ have the same bids. Thus $v_i \succcurlyeq v_{i+1}$ and $v_{i+1} \succcurlyeq v_i$. So, $r_i = f(v_1, v_2, \ldots, v_{i-1}, v_{i+1}, \ldots, v_n)$ and $r_{i+1} = f(v_1, v_2, \ldots, v_i, v_{i+2}, \ldots, v_n)$. Since $v_i = v_{i+1}$, $r_i = r_{i+1}$. Thus the rebate function is anonymous.

- The "only if" part: For the mechanism to be strategyproof, the rebate function for agent $i$ should be independent of his bid. So, $r_i$ should depend on only $v_{-i}$. So, for deterministic rebate function, $r_i = f_i(v_{-i})$. Now, we desire anonymous rebate function. That is, rebate should be independent of the identity of the agent. Thus, if $v_i = v_j$, then $r_i = r_j$. With out loss of generality, say $v_i = v_{i+1}$, then $v_{-i} = v_{-(i+1)}$. So, $r_i = r_{i+1}$ implies, $f_i = f_{i+1}$. Similarly $f_{i+1} = f_{i+2}$ and so on. Thus, $r_i = f(v_{-i}) \ \forall i \in N$.

$\square$

We now state and prove the main result of this paper.

**Theorem 3** *If a redistribution mechanism is feasible and individually rational, then there cannot exist a linear rebate function which simultaneously satisfies all the following properties:*

- *DSIC*

- *deterministic*

- *anonymous*

- *non-zero redistribution index.*

**Proof :** Assume to the contrary that there exists a linear function, say $f$, which satisfies the above properties. Let $v_1 \succcurlyeq v_2 \succcurlyeq \ldots \succcurlyeq v_n$. Then according to Theorem 2, for each agent $i$,

$$
\begin{aligned}
r_i &= f(v_1, v_2, \ldots, v_{i-1}, v_{i+1}, \ldots, v_n) \\
&= (c_0, e_p) + (c_1, v_1) + \ldots + (c_{n-1}, v_n)
\end{aligned}
$$

where, $c_i = (c_{i1}, c_{i2}, \ldots, c_{ip}) \in \mathbb{R}^p$, $e_p = (1, 1, \ldots, 1) \in \mathbb{R}^p$, and $(\cdot, \cdot)$ denotes the inner product of two vectors in $\mathbb{R}^p$. Now, we will show that the worst case performance of $f$ will be zero. To this end, we will study the structure of $f$, step by step.

<u>Observation 1:</u> Consider type profile $(v_1, v_2, \ldots, v_n)$ where $v_1 = v_2 = \ldots = v_n = (0, 0, \ldots, 0)$. For this type profile, the total Clarke surplus is zero and $r_i = (c_0, e_p) \ \forall i \in N$. Individual rationality implies,

$$(c_0, e_p) \geq 0 \tag{5}$$

Feasibility implies the total redistributed amount is less than the surplus, that is,

$$\sum_i r_i = n(c_0, e_p) \leqslant 0 \tag{6}$$

From, (5) and (6), it is easy to see that, $(c_0, e_p) = 0$.

<u>Observation 2:</u> Consider type profile $(v_1, v_2, \ldots, v_n)$ where $v_1 = (1, 0, 0, \ldots, 0)$ and $v_2 = \ldots, v_n = (0, 0, \ldots, 0)$. For this type profile, $r_1 = 0$ and if $i \neq 1$, $r_i = c_{11} \geq 0$ for individual rationality. For this type profile, it can be seen through straight forward calculations that the Clarke surplus is zero. Thus, for feasibility, $\sum_i r_i = (n-1)c_{11} \leq t = 0$. This implies, $c_{11} = 0$.

In the above profile, by considering $v_1 = (0, 1, 0, \ldots, 0)$, we get $c_{12} = 0$. Similarly, one can show $c_{13} = c_{14} = \ldots = c_{1p} = 0$.

<u>Observation 3:</u> Continuing on the same lines as above with, $v_1 = v_2 = \ldots = v_i = e_p$, and $v_{i+1} = (1, 0 \ldots, 0)$
or $(0, 1, 0 \ldots, 0), \ldots$ or $(0, \ldots, 0, 1)$, we get, $c_{i+1} = (0, 0, \ldots, 0) \ \forall \ i \leq p-1$. Thus,

$$
r_i = \begin{cases}
(c_{p+1}, v_{p+2}) + \ldots + (c_{n-1}, v_n) & : \text{if } i \leq p+1 \\
(c_{p+1}, v_{p+1}) + \ldots + (c_{i-1}, v_{i-1}) & \\
\quad + (c_i, v_{i+1}) + \ldots + (c_{n-1}, v_n) & : \text{otherwise}
\end{cases} \tag{7}
$$

Thus a rebate function in any linear redistribution mechanism has to be necessarily of the form in the Equation (7). We now claim that the redistribution index of such a mechanism is zero. For any individually rational redistribution mechanism, the trivial lower bound on redistribution index is zero. We prove that in a linear redistribution mechanism, there exists a type profile, at which the fraction of the Clarke surplus that gets redistributed is zero. Consider the type profile:

$$
\begin{aligned}
v_1 &= (2p-1, 2p-2, \ldots, p+1, p) \\
v_2 &= (2p-2, 2p-3, \ldots, p, p-1) \\
&\vdots \\
v_{p-1} &= (p+1, p, \ldots, 3, 2) \\
v_p &= (p, p-1, \ldots, 2, 1)
\end{aligned}
$$

and $v_{p+1} = v_{p+2} \ldots = v_n = (0, 0, \ldots, 0)$.

Now it can be seen, through straightforward calculations of the Clarke payments, that, with this type profile, agent 1 pays $(p-1)$, agent 2 pays $(p-2), \ldots,$ agent $(p-1)$ pays 1 and the remaining agents pay 0. Thus, the Clarke payment received is non-zero but it can be seen that $r_i = 0$ for all the agents. Hence, the redistribution index for any linear redistribution mechanism has to be zero.

<div align="right">□</div>

The above theorem provides a disappointing piece of news. It rules out the possibility of a linear redistribution mechanism for heterogeneous settings which will have non-zero redistribution index. However, there are two ways to get around it.

1. The domain of types under which Theorem 3 holds is, $\Theta_i = \mathbb{R}^p_+, \forall i \in N$. One idea is to restrict the domain of types. In Section 4, we design a worst case optimal linear redistribution mechanism when the valuations of agents for the heterogeneous objects have a certain type of relationship.

2. Explore the existence of a rebate function which is not linear but yields a non-zero redistribution index. We explore this in Section 5.

It should be noted that our impossibility result holds true when we are defining a linear rebate functions as in Definition 4. Our result may not hold for other types of linearity. For example, sort bid components of other $(n-1)$ agents and define rebate function to be linear combination of these $(n-1)p$ elements. At this point, we have not explored such linear rebate functions.

## 4. A Redistribution Mechanism for Heterogeneous Objects when Valuations have a Scaling Based Relationship

Consider a scenario where the objects are not identical but the valuations for the objects are related and can be derived by a single parameter. As a motivating example, consider there is a website where people can put up their ads for free and assume that there are $p$ slots available for advertisements and there are $n$ agents interested in displaying their ads. Naturally, every agent will have a higher preference for a higher slot. Another motivating example could be, there is university web site which has $p$ slots to display news about various departments. Define *click through rate* of a slot as the number of times the ad is clicked, when the ad is displayed in that slot, divided by the number of impressions. Let the click through rates for slots be $\gamma_1 \geq \gamma_2 \geq \gamma_3 \ldots \geq \gamma_p$. Assume that each agent has the same value for each click by the user, say $v_i$. So, the agent's value for the $j^{th}$ slot will be $\gamma_j v_i$. Let us use the phrase *valuations with scaling based relationship* to describe such valuations. We define this more formally below.

**Definition 6** *We say the valuations of the agents have scaling based relationship if there exist positive real numbers $\gamma_1, \gamma_2, \gamma_3, \ldots, \gamma_p > 0$ such that, for each agent $i \in N$, the valuation for object $j$, say $\theta_{ij}$, is of the form $\theta_{ij} = \gamma_j v_i$, where $v_i \in \mathbb{R}_+$ is a private signal observed by agent $i$.*

Without loss of generality, we assume, $\gamma_1 \geq \gamma_2 \geq \gamma_3 \ldots \geq \gamma_p > 0$. (For simplifying equations, we will assume that there are $(n - p)$ virtual objects, with $\gamma_{p+1} = \gamma_{p+2} = \ldots = \gamma_n = 0$). We immediately note that the homogeneous setting is a special case that arises when $\gamma_1 = \gamma_2 = \gamma_3 = \ldots = \gamma_p > 0$

For the above setting, we design a Groves mechanism which is almost budget balanced and optimal in the worst case. Our mechanism is similar to that of Guo and Conitzer (2009) and our proof uses the same line of arguments.

## 4.1 The Proposed Mechanism

We will use a linear rebate function. We propose the following mechanism:

- The agents submit their bids.

- The bids are sorted in decreasing order.

- The highest bidder will be allotted the first object, the second highest bidder will be allotted the second object, and so on.

- Agent $i$ will pay $t_i - r_i$, where $t_i$ is the Clarke payment and $r_i$ is the rebate.

$$t_i = \sum_{j=i}^{p} (\gamma_j - \gamma_{j+1}) v_{j+1}$$

- Let agent $i$'s rebate be,

$$r_i = c_0 + c_1 v_1 + \ldots + c_{i-1} v_{i-1} + c_i v_{i+1} + \ldots + c_{n-1} v_n$$

  $c_i$'s are defined as follows.

The mechanism is required to be individually rational and feasible.

- The mechanism will be individually rational iff $r_i \geq 0 \; \forall i \in N$. That is, $\forall i \in N$,

$$c_0 + c_1 v_1 + \ldots + c_{i-1} v_{i-1} + c_i v_{i+1} + \ldots + c_{n-1} v_n \geq 0.$$

- The mechanism will be feasible if the total redistributed payment is less than or equal to the surplus. That is, $\sum_i r_i \leq t = \sum_i t_i$ or $t - \sum_i r_i \geq 0$, where,

$$t = \sum_{j=1}^{p} j(\gamma_j - \gamma_{j+1})v_{j+1}.$$

With the above setup, we now derive $c_0, c_1, \ldots, c_{n-1}$ that will maximize the fraction of the surplus which is redistributed among the agents.

Step 1: First, we claim that $c_0 = c_1 = 0$. This can be proved as follows. Consider the type profile, $v_1 = v_2 = \ldots = v_n = 0$. For this type profile, individual rationality implies $r_i = c_0 \geq 0$ and $t = 0$. So for feasibility, $\sum_i r_i = nc_0 \leq t = 0$. That is, $c_0$ should be zero. Similarly, by considering type profile $v_1 = 1, v_2 = \ldots = v_n = 0$, we get $c_1 = 0$.

Step 2: Using $c_0 = c_1 = 0$,

- The feasibility condition can be written as:

$$\sum_{j=2}^{n-1} \left\{ (j-1)(\gamma_{j-1} - \gamma_j) - (j-1)c_{j-1} - (n-j)c_j \right\} v_j - (n-1)c_{n-1}v_n \geq 0 \qquad (8)$$

- The individual rationality condition can be written as

$$c_2 v_2 + \ldots + c_{i-1}v_{i-1} + c_i v_{i+1} + \ldots + c_{n-1}v_n \geq 0 \qquad (9)$$

Step 3: When we say our mechanism's redistribution index is $e$, we mean, $\sum_i r_i \geq et$, that is,

$$\sum_{j=2}^{n-1} \Big( -e(j-1)(\gamma_{j-1} - \gamma_j) + (j-1)c_{j-1} + (n-j)c_j \Big) v_j +$$
$$(n-1)c_{n-1}v_n \geq 0 \qquad (10)$$

Step 4: Define $\beta_1 = \gamma_1 - \gamma_2$, and for $i = 2, \ldots, n-1$, let $\beta_i = i(\gamma_i - \gamma_{i+1}) + \beta_{i-1}$. Now, inequalities (8), (9), and (10) have to be satisfied for all values of $v_1 \geq v_2 \geq \ldots \geq v_n \geq 0$. By Theorem (1), we need to satisfy the following set of inequalities:

$$\sum_{i=2}^{j} c_i \geq 0 \ \ \forall j = 2, \ldots n-1$$
$$e\beta_1 \leq (n-2)c_2 \leq \beta_1$$
$$e\beta_{i-1} \leq n \sum_{j=2}^{i-1} c_j + (n-i)c_i \leq \beta_{i-1} \ \ i = 3, \ldots, p$$
$$e\beta_p \leq n \sum_{j=2}^{i-1} c_j + (n-i)c_i \leq \beta_p \ \ i = p+1, \ldots, n-1$$
$$e\beta_p \leq n \sum_{j=2}^{n-1} c_j \leq \beta_p$$

Now, the mechanism designer wishes to design a mechanism that maximizes $e$ subject to the above constraints.

Define $x_j = \sum_{i=2}^{j} c_i$ for $j = 2, \ldots, n-1$. This is equivalent to solving the following linear program.

$$
\boxed{
\begin{array}{c}
\text{maximize } e \\
s.t. \\
e\beta_1 \leq (n-2)x_2 \leq \beta_1 \\
e\beta_{i-1} \leq ix_{i-1} + (n-i)x_i \leq \beta_{i-1} \ \ i = 3, \ldots, p \\
e\beta_p \leq ix_{i-1} + (n-i)x_i \leq \beta_p \ \ i = p+1, \ldots, n-1 \\
e\beta_p \leq nx_{n-1} \leq \beta_p \\
x_i \geq 0 \ \ \forall i = 2, \ldots, n-1
\end{array}
}
\tag{11}
$$

$\square$

So, given $n$ and $p$, the social planner will have to solve the above optimization problem and determine the optimal values of $e, c_2, c_3, \ldots, c_{n-1}$. It would be of interest to derive a closed form solution for the above problem.

The discussion above can be summarized as the following theorem.

**Theorem 4** *When the valuations of the agents have scaling based relationship, for any $p$ and $n > p+1$, the linear redistribution mechanism obtained by solving LP (11) worst case optimal among all Groves redistribution mechanisms that are feasible, individually rational, deterministic, and anonymous. This mechanism is an example of a mechanism having non-zero redistribution index.*

**Proof:**
The worst case optimality of the mechanism can be proved following the line of arguments of Guo and Conitzer (2009).

As per the impossibility Theorem 3, there is no linear redistribution mechanism for general heterogeneous setting having non-zero efficiency. However, when objects have scaling based relationship, the linear redistribution mechanism, that is obtained by solving LP (11) has non-zero efficiency at least for some $(n, p)$ instances. This is obtained by actually solving the LP (for example, using MATLAB) for various values of $n$ and $p$. This certainly proves that, at least for $n = 10, 12, 14, p = 2, 3, 4, \ldots, 8$ and when valuations have scaling based correlation, the worst case optimal mechanism given by the LP (11) has non-zero redistribution index. Now we obtain an upper bound on the redistribution index of a redistribution mechanism LP (11).

**Claim 1** *If $e^*$ is the solution of the LP (11), then*

$$
e^* \leq \min\left\{\frac{A}{B}, \frac{B}{A}\right\}
$$

*where, $A = \sum_{i=1,3,5,\ldots} \beta_{i-1} \begin{pmatrix} n \\ i \end{pmatrix}$ and $B = \sum_{i=2,4,6,\ldots} \beta_{i-1} \begin{pmatrix} n \\ i \end{pmatrix}$.*

The LP (11) can be written as

$$
\boxed{
\begin{array}{c}
\text{maximize } e \\
s.t. \\
e\beta \leq Mx \leq \beta \\
x \geq 0
\end{array}
}
$$

where $x = (x_2, x_3, \ldots, x_{n-1}) \in \mathbb{R}_+^{n-2}$ and $\beta = (\beta_1, \beta_2, \ldots, \beta_p, \beta_p, \ldots, \beta_p) \in \mathbb{R}_+^{n-1}$ and

$$M = \begin{bmatrix} n-2 & 0 & 0 & \cdots & 0 \\ 3 & n-3 & 0 & \cdots & 0 \\ 0 & 4 & n-4 & \cdots & 0 \\ \vdots & \vdots & \vdots & \vdots & \vdots \\ 0 & & & n-1 & 1 \\ 0 & & & 0 & n \end{bmatrix}$$

Now, $y = (y_1, y_2, \ldots, y_{n-1}) \in \mathbb{R}^{n-1}$ is in the range of $M$ *iff*

$$\binom{n}{2} y_1 + \binom{n}{4} y_3 + \cdots = \binom{n}{3} y_2 + \binom{n}{5} y_4 + \cdots \tag{12}$$

Now, $e^* \beta_i \binom{n}{i+1} \leq \binom{n}{i+1} (Mx)_i \leq \binom{n}{i+1} \beta_i \; \forall i \in \{1, 2, 3, \ldots, n-1\}$. Now summing these inequalities for odd $i$s and using (12), we get $e^* A \leq B$ and summing over even $i$s we get $e^* B \leq A$. This proves our claim.

$\square$

We verified using MATLAB for $n = 10, 12, 14$ and $p = 2, 3, \ldots 8$, that the redistribution index of the proposed mechanism is in fact, $e^* = \min\left\{\frac{A}{B}, \frac{B}{A}\right\}$.

## 5. Non-linear Redistribution Mechanisms for the Heterogeneous Setting

We should note that the homogeneous objects case is a special case of the heterogeneous objects case in which each bidder submits the same bid for all objects. Thus, we cannot expect any redistribution mechanism to perform better than the homogeneous objects case. For $n \leq p+1$, the worst case redistribution is zero for the homogeneous case and so will it be for the heterogeneous case (Guo & Conitzer, 2009; Moulin, 2009). So, we assume $n > p+1$. In this section, we propose two redistribution mechanisms with non-linear rebate functions. We construct a redistribution scheme by applying the mechanism proposed by Bailey (1997) to the heterogeneous settings. We refer to this proposed mechanism on heterogeneous objects as *BAILEY-CAVALLO* redistribution mechanism. It is crucial to note that the non-zero redistribution index of the BAILEY-CAVALLO mechanism does not trivially follow from that of the mechanism in the work of Bailey. We rewrite the WCO mechanism and extend the rebate functions to heterogeneous objects settings. We call this mechanism as HETERO.

In each of the mechanisms, namely BAILEY-CAVALLO and HETERO, the objects are assigned to those agents who value them most. The Clarke payments are collected from the agents and the surplus is redistributed among the agents according to the rebate functions defined in the mechanism. Hence, both are Groves redistribution mechanisms and hence DSIC.

As stated above, for $n \leq (p+1)$, the redistribution index for any redistribution mechanism has to be zero. For the case $n > p+1$, the redistribution index for any linear redistribution mechanism has to be zero (Theorem 3). We prove, for $n \geq (2p+1)$, that BAILEY-CAVALLO has non-zero redistribution index. We only conjecture that HETERO is worst case optimal, that is no mechanism can have better redistribution index than HETERO. We also conjecture that HETERO's redistribution index is the same as that of WCO, which is non-zero when $n > (p + 1)$. Thus, for $n \in \{p + 2, p + 3, \ldots, 2p\}$, there is still no redistribution mechanism for which non-zero redistribution index is proved.

## 5.1 BAILEY-CAVALLO Mechanism

First, consider the case when $p = 1$. Let the valuations of the agents for the object be, $v_1 \geq v_2 \geq \ldots \geq v_n$. The agent with the highest valuation will receive the object and would pay the second highest bid. Cavallo (2006) proposed the rebate function as:

$$r_1 = r_2 = \frac{1}{n}v_3$$
$$r_i = \frac{1}{n}v_2 \quad i > 2$$

A similar mechanism was independently proposed by Porter et al (2004). Motivated by this scheme, we propose a scheme for the heterogeneous setting. Suppose agent $i$ is excluded from the system. Then let $t^{-i}$ be the Clarke surplus in the system (defined in Table 1). Define,

$$r_i^{\ B} = \frac{1}{n}t^{-i} \ \ \forall i \in N \tag{13}$$

- As the Clarke surplus is always positive, $r_i^{\ B} \geq 0$ for all $i$. Thus, this scheme satisfies individual rationality.

- $t^{-i} \leq t \quad \forall \ i$ (revenue monotonicity). So, $\sum_i r_i^{\ B} = \sum_i \frac{1}{n}t^{-i} \leq n\frac{1}{n}t = t$. Thus, this scheme is feasible. (The revenue monotonicity follows from the fact that the valuations are non-negative and unit demand preferences. Gul and Stacchetti (1999) showed that with unit demand preferences, the VCG payments coinside with the smallest Walrasian prices which in turn would not decrease by addition of an agent. Thus addition of any agent cannot decrease the total payments.)[2]

We now show that the BAILEY-CAVALLO scheme has non-zero redistribution index if $n \geq 2p + 1$. First we state two lemmas. The proof will be given in Appendix B. These lemmas are useful in designing redistribution mechanisms for the heterogeneous settings as well as in analysis of the mechanisms. Lemma 2 is used to show that the redistribution index of the BAILEY-CAVALLO mechanism is non-zero. Lemma 1 is used to find an allocatively efficient outcome for the settings under consideration. Lemma 1 is also useful in determining the Clarke payments.

**Lemma 1** *If we sort the bids of all the agents for each object, then:*

1. *An optimal allocation, that is an allocatively efficient allocation, will consist of the agents having bids among the $p$ highest bids for each object.*

2. *Consider an optimal allocation $k^*$. If any of the $p$ agents receiving objects in $k^*$ is dropped, then there always exists an allocation $k^*_{-i}$ that is an optimal allocation (on*

---

2. We thank the annonymous reviewer for pointing out this reference.

*the remaining $n-1$ agents) which allocates objects to the remaining $(p-1)$ agents. The objects that these $(p-1)$ agents receive in $k^*_{-i}$, may not however be the same as the objects they are allocated in $k^*$.*

**Lemma 2** *There are at most $2p$ agents involved in deciding the Clarke payment.*

*Note*: When the objects are identical, the bids of $(p+1)$ agents are involved in determining the Clarke payments.

Now, we show that the redistribution index of the BAILEY-CAVALLO mechanism is non-zero.

**Theorem 5** *When there are sufficient number of agents (in particular, $n > 2p$), the BAILEY-CAVALLO redistribution mechanism has non-zero redistribution index.*

**Proof:** In Lemma 2, we have shown that there will be at most $2p$ agents involved in determining the Clarke surplus. Thus, given a type profile, there will be $(n-2p)$ agents, for whom, $t^{-i} = t$ and this implies that at least $\frac{n-2p}{n}t$ will be redistributed. That is the redistribution index of the mechanism is at least $\frac{n-2p}{n} > 0$.

$\square$

Note that the above mechanism may not be worst case optimal. This is because, when objects are identical, the WCO mechanism performs better on worst case analysis than the above mechanism. So, we suspect that in heterogeneous settings as well, the above mechanism would not be optimal on worst case analysis. In the next subsection, we explore another rebate function, namely HETERO.

## 5.2 HETERO: A Redistribution Mechanism for the Heterogeneous Setting

When the objects are identical, the WCO mechanism is given by equation (3). We give a novel interpretation to it. Consider the scenario in which one agent is absent from the scene. Then the Clarke payment received is either $pv_{p+1}$ or $pv_{p+2}$ depending upon which agent is absent. If we remove two agents, the surplus is $pv_{p+1}$ or $pv_{p+2}$ or $pv_{p+3}$, depending upon which two agents are removed. Till $(n-p-1)$ agents are removed, we get non-zero surplus. If we remove $(n-p)$ or more agents from the system, there is no need for any mechanism for assignment of the objects. So, we will consider the cases when we remove $k$ agents, where, $1 \le k < n-p$.

Now let $t^{-i,k}$ be the average payment received when agent $i$ is removed along with $k$ other agents that is, a total of $(k+1)$ agents are removed comprising of $i$. The average is taken over all possible selections of $k$ agents from the remaining $(n-1)$ agents. We can rewrite the WCO mechanism in terms of $t^{-i}, t^{-i,k}$. Observe that, $t^{-i}, t^{-i,k}$ can be defined in heterogeneous settings as well. We propose to use a rebate function defined as,

$$r_i^H = \alpha_1 t^{-i} + \sum_{k=2}^{k=n-p-1} \alpha_k t^{-i,k-1} \tag{14}$$

where $\alpha_k$ are the suitable weights assigned to the surplus generated when a total of $k$ agents are removed from the system. By using different $\alpha_k$s, we get different mechanisms. However, we prefer to choose $\alpha_k$s as the following.

145

### 5.2.1 THE EQUIVALENCE OF HETERO AND WCO WHEN OBJECTS ARE IDENTICAL

It is desirable that HETERO should match with the WCO mechanism when the objects are homogeneous. So we choose $\alpha$'s in Equation (14) in a way that ensures that, when the objects are identical, $r_i^H$ in equation (14) is equal to $r_i^{WCO}$ in equation (3) for all type profiles. Since the rebate is a function of the remaining $(n-1)$ bids, we can write it as, $r_i = f(x_1, x_2, \ldots, x_{n-1})$ where $x_1, x_2, \ldots, x_{n-1}$ are the bids without the agent $i$, in decreasing order. Note, in this case, that each bidder will be submitting a bid $b_i \in \mathbb{R}_+$.

Now, we can write, $t^{-i,k}, r_i^H$, and $r_i$ in terms of $x_1, x_2, \ldots, x_{n-1}$, as,

$$t^{-i,k-1} = \sum_{l=0}^{k-1} \frac{\binom{p+l}{p}\binom{n-p-2-l}{k-1-l}}{\binom{n-1}{k-1}} x_{p+1+l}$$

$$r_i^H = \sum_{k=1}^{k=n-p-1} \alpha_k \, t^{-i,k-1} \tag{15}$$

$$r_i^{WCO} = \sum_{l=0}^{n-p-1} c_{p+1+l} \, x_{p+1+l} \tag{16}$$

where, $c_i, \; i = p+1, p+2, \ldots, n-1$ are given by equation (2).

Consider the type profile $(x_1 = 1, x_2 = 1, \ldots, x_{p+1} = 1, x_{p+2} = 0, \ldots, x_{n-1} = 0)$. For HETERO to agree with WCO, the coefficients of $x_{p+1}$ in equation (15) and equation (16) should be the same. Now consider the type profile $(x_1 = 1, x_2 = 1, \ldots, x_{p+2} = 1, x_{p+3} = 0, \ldots, x_{n-1} = 0)$. As the coefficients of $x_{p+1}$ in equation (15) and equation (16) are the same, the coefficients of $x_{p+2}$ should also be equal in equation (15) and equation (16).

Thus, the coefficients of $x_{p+1}, x_{p+2}, \ldots, x_{n-1}$ in equation (15) and equation (16) should agree.

Let $L = n - p - 1$. Thus, for $i = p+1, \ldots, n-1$,

$$c_i = \sum_{k=0}^{n-i-1} \alpha_{L-k} \frac{\binom{i-1}{p}\binom{n-i-1}{k}}{\binom{n-1}{p+1+k}} \tag{17}$$

The above system of equations yields, for $i = 1, 2, \ldots, L$,

$$\alpha_i = \frac{(-1)^{(i+1)}(L-i)!\,p!}{(n-i)!} \chi \sum_{j=0}^{L-i} \left\{ \binom{i+j-1}{j} \sum_{l=p+i+j}^{n-1} \binom{n-1}{l} \right\} \tag{18}$$

where $\chi$ is given by, $\chi = \dfrac{(n-p)\binom{n-1}{p-1}}{\sum_{j=p}^{n-1}\binom{n-1}{j}}$.

### 5.2.2 Properties of HETERO

As the HETERO mechanism matches with the WCO when objects are identical, the HETERO mechanism satisfies individual rationality and feasibility in the homogeneous case. These two properties, however, remain to be shown in the heterogeneous case.

**Conjecture 1** *The HETERO mechanism satisfies individual rationality, feasibility, is worst case optimal, and has redistribution index same as WCO.*

### 5.2.3 Intuition Behind Individual Rationality of HETERO

We have to show that for each agent $i$, $r_i^H \geq 0$ at all type profiles. For convenience, we will assume $i$ implicitly. So, say, $r_i^H = r$ and $\Gamma_1 = t^{-i}$, $\Gamma_j = t^{-i,j-1}$, $j = 2, \ldots, L$. Now, the rebate is given by the equation, $r = \sum_j \alpha_j \Gamma_j$. We have to show that $r \geq 0$. Note that, $\Gamma_1 \geq \Gamma_2 \geq \ldots \geq \Gamma_L \geq 0$. The $\Gamma$'s are monotone as the absence of more agents would either decrease the VCG payments or the payments remain the same. So, if $\sum_{i=1}^{j} \alpha_i \geq 0 \ \forall \ j = 1 \rightarrow L$, individual rationality would follow from Theorem 1. We observe that, in general, this is not true. The important observation is, though $\Gamma_i$'s are decreasing positive real numbers, they are related. For example, we can show that if $\Gamma_1 > 0$, then $\Gamma_2 > 0$. In our experiments, which we describe in the next section, we keep track of $\frac{\Gamma_2}{\Gamma_1}$. We observed that this ratio is in $[0.5, 1]$. For Theorem 1 to be applicable, this ratio can be any value in $[0, 1]$.

Thus, though $\alpha$'s are alternately positive and negative, the relation among $\Gamma$'s would not make $r$ to become negative and it will be within limits in such a way that total rebate to the agents will be less than or equal to total Clarke payment. It remains to show individual rationality analytically in the general case. We are, however, only able to show in the following cases.

1. Consider the case when $p = 2$. (i). If $n = 4$, $\alpha_1 = \frac{1}{4}$. (ii). If $n = 5$, $\alpha_1 = 0.27273$, $\alpha_2 = -0.18182$. (iii). If $n = 6$, $\alpha_1 = 0.29487$, $\alpha_2 = -0.25641$, $\alpha_3 = 0.12821$.

2. Consider the case when $p = 3$. (i). If $n = 5$, $\alpha_1 = \frac{1}{5}$. (ii). If $n = 6$, $\alpha_1 = 0.21875$, $\alpha_2 = -0.15625$. (iii). If $n = 7$, $\alpha_1 = 0.23810$, $\alpha_2 = -0.21429$, $\alpha_3 = 0.11905$.

By Theorem 1, it follows that for the above cases, the proposed mechanism satisfies individual rationality.

### 5.2.4 Feasibility and Worst Case Optimality of HETERO

Similarly, we also believe that, $\alpha$'s adjust the rebate functions optimally such that, HETERO remains feasible and is worst case optimal and has the same redistribution index as WCO. Though we do not have analytical proof, we provide some empirical evidence for the conjecture in Section 6.

## 6. Experimental Analysis

We perform our experiments in two sets. In the first set, we consider bids to be real numbers. In the second set we consider the bidders submitting binary bids. We use these experiments to provide an empirical evidance for our Conjecture 1.

## 6.1 Empirical Evidence for Individual Rationality of HETERO

Solving equations (18) is a challenging task. Though the new mechanism is the extension of the Moulin or the WCO mechanism, yet, we are not able to prove individual rationality and feasibility of HETERO analytically. We therefore seek empirical evidence.

### 6.1.1 Simulation 1

We consider various combinations of $n$ and $p$. For each agent, and for each object, the valuation is generated as a uniform random variable in $[0, 100]$. We run our simulations for the following combinations of $n$ and $p$.

For $p = 2$, $n = 5, 6, \ldots, 14$, for $p = 3$, $n = 7, 8, \ldots, 14$ and for $p = 4$, $n = 9, 10, \ldots, 14$. For each combination of $n$ and $p = 2$, we generated randomly 100,000 bid profiles and evaluated our mechanism. We also kept track of the worst case performance of our mechanism over these 100,000 bid profiles. Our mechanism was feasible and individually rational in these 100,000 bid profiles. The redistribution index of our mechanism is upper bounded by that of the WCO mechanism. We observed that the worst case performance over these 100,000 random bid profiles was the same as that of WCO. This is a strong indication that our mechanism will perform well in general.

### 6.1.2 Simulation 2: Bidders with Binary Valuation

Suppose each bidder has valuation for each object, either 0 or 1. Then there are $2^{np}$ possible bid profiles. We ran an experiment to evaluate our mechanism with all possible bid profiles of agents with binary valuations. We considered $p = 2$ and $n = 5, 6, \ldots, 12$. We found that the mechanism is feasible, individually rational, and the worst case performance is the same as that of the WCO mechanism. Note, as indicated earlier, no mechanism can perform better than the WCO mechanism in the worst case. And our mechanism performs equally well as the WCO. Thus, though an analytical proof is elusive, for binary valuation settings, for $p = 2$ and $n = 5, 6, \ldots, 12$, our mechanism is worst case optimal.

## 6.2 BAILEY-CAVALLO vs HETERO

In this subsection, we compare the worst case redistribution index of BAILEY-CAVALLO with the worst case redistribution index of HETERO for varying number of objects when there are 10 agents in the system. That is, we study worst case redistribution index for various $p$ when $n = 10$. The worst case is taken over randomly generated 50K bid profiles. The comparison is depicted in Figure 1. The redistribution index of WCO is an upper bound on any Redistribution Mechanism for heterogeneous settings. However, the simulations not being exhaustive, the worst case performance of the mechanisms could perhaps be better than that of WCO. Exact worst case may be worse than WCO. However, in the simulations, we never encountered a situation where HETERO is worse than WCO. We can see from Figure 1 that BAILEY-CAVALLO mechanism's worst case performance is better than that of HETERO, for $p = 3, 4, 5, 6, 7$. This worst case is the worst over 50,000 randomly generated bid profiles in our simulations.

The other observation we made in our simulations is that most of the time (70%), BAILEY-CAVALLO redistributes more VCG surplus than HETERO ever though the worst case performance is worse than that of HETERO.

These observations also lead to a question that Cavallo (2008) raised in the context of dynamic redistribution mechanisms. Do we really need a highly sophisticated mechanism, that is worst case optimal, when a simple mechanism performs quite well in general.

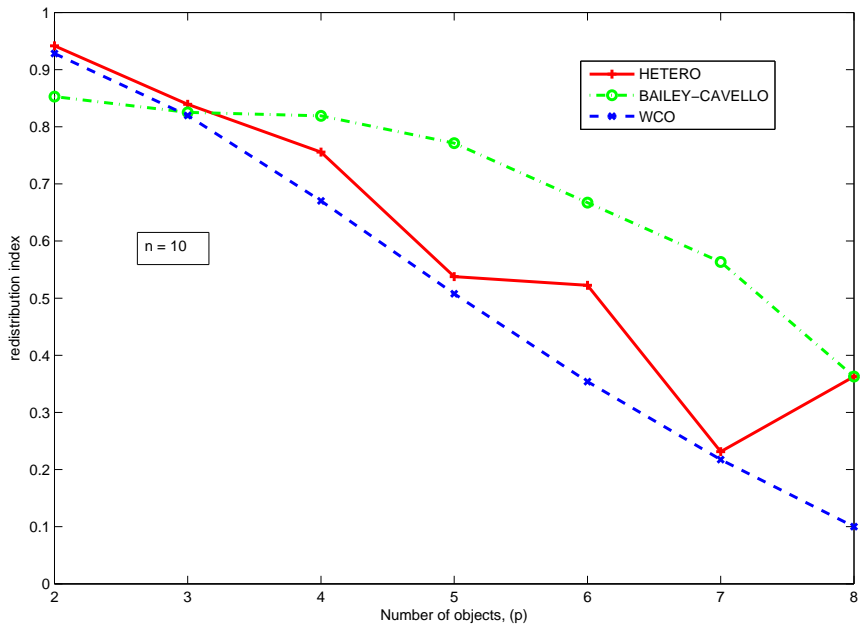

Figure 1: Redistribution index vs number of objects when number of agents = 10

## 7. Conclusion

We addressed the problem of assigning $p$ heterogeneous objects among $n > p$ competing agents. When the valuations of the agents are independent of each other and their valuations for each object are independent of valuations on the other objects, we proved the impossibility of existence of a linear redistribution mechanism with non-zero redistribution index (Theorem 3). Then we explored two approaches to get around this impossibility.

- In the first approach, we showed that linear rebate functions with non-zero redistribution index are possible when the valuations for the objects have a scaling based relationship. For these settings, we proposed a strategyproof linear redistribution mechanism that is optimal on worst case analysis, individually rational, and feasible (Theorem 4).

- In the second approach, we relaxed linearity requirement. We showed that non-linear rebate functions with non-zero redistribution index are possible by applying the BAILEY-CAVALLO mechanism to the settings (Theorem 5).

- We proposed a mechanism, namely HETERO, for general settings when the objects are heterogeneous and the private values of an agent for these objects are independent of each other. The mechanism is deterministic, anonymous, and DSIC. The HETERO mechanism extends the Moulin /WCO mechanism. Though we have not analytically proved feasibility and individual rationality, we have sufficient empirical evidence to conjecture that our mechanism is feasible and individually rational (Conjecture 1).

It would be interesting to see if we can characterize the situations under which linear redistribution mechanisms with non-zero redistribution indices are possible for heterogeneous settings.

An interesting research direction is to investigate the individual rationality and feasibility for the proposed HETERO mechanism. Also, we strongly believe that this mechanism is a worst case optimal. An immediate future direction is to prove this fact or design a mechanism which is worst case optimal.

Another interesting problem to explore is to characterize all redistribution mechanisms that are worst case optimal in heterogeneous settings.

## Acknowledgments

The first author would like to acknowledge Infosys Technologies Ltd., for awarding Infosys fellowship to pursue Ph.D. The authors would like to thank Professor David Parkes for useful comments. The authors would also like to thank the anonymous reviewers, whose feedback has helped a lot in improving this paper.

## Appendix A. Ordering of the Agents Based on Bid Profiles

We will define a ranking among the agents. This ranking is used in proving a Theorem 2 on rebate function. This theorem is similar to Cavallo's theorem on characterization of DSIC, deterministic, anonymous rebate functions for homogeneous objects. We would not be actually computing the order among the bidders. We will use this order for proving impossibility of the linear rebate function with the desired properties.

### A.1 Properties of the Ranking System

When we are defining ranking/ordering among the agents, we expect the following properties to hold true:

- Any permutation of the objects and the corresponding permutation on bid vector,$(b_{i1}, b_{i2}, \ldots, b_{ip})$ for each agent $i$, should not change the ranking. That is, the ranking should be independent of the order in which the agents are expected to bid for this objects.

- Two bidders with the same bid vectors should have the same rank.

- By increasing the bid on any of the objects, the rank of an agent should not decrease.

## A.2 Ranking among the Agents

This is a very crucial step. First, find out all feasible allocations of the $p$ objects among the $n$ agents, each agent receiving at most one object. Sort these allocations, according to the valuation of an allocation. Call this list $\mathcal{L}$. To find the ranking between $i$ and $j$, we uses the following algorithm.

1. $\mathcal{L}_{ij} = \mathcal{L}$

2. Delete all the allocations from $\mathcal{L}_{ij}$ which contain both $i$ and $j$.

3. Find out the first allocation in $\mathcal{L}_{ij}$ which contains one of the agent $i$ or $j$. Say $k'$.

   (a) Suppose this allocation contains $i$ and has value strictly greater than any of remaining allocations from $\mathcal{L}_{ij}$ containing $j$, then we say, $i \succ j$.

   (b) Suppose this allocation contains $j$ and has value strictly greater than any of remaining allocations from $\mathcal{L}_{ij}$ containing $i$, then we say, $j \succ i$.

4. If the above step is not able to decide the ordering between $i$ and $j$, let $\mathcal{A} = \{k \in K | v(k) = v(k')\}$. Update $\mathcal{L}_{ij} = \mathcal{L}_{ij} \setminus \mathcal{A}$ and recur to step (2) till EITHER
   - there is no allocation containing the agent $i$ or $j$ OR
   - the ordering between $i$ and $j$ is decided.

5. If the above steps do not give either of $i \succ j$ or $j \succ i$, we say, $i \equiv j$ or $i \succcurlyeq j$ as well as $j \succcurlyeq i$.

Before we state some properties of this ranking system $\succcurlyeq$, we will explain it with an example. Let there be two items A and B, and four bidders. That is, $p = 2, n = 4$ and let their bids be: $b_1 = (4, 5), b_2 = (2, 1), b_3 = (1, 4),$ and $b_4 = (1, 0)$.

Now, allocation $(A = 1, B = 3)$ has the highest valuation among all the allocations. So,

$$\text{agent } 1 \succ \text{ agent } 2$$
$$\text{agent } 1 \succ \text{ agent } 4$$
$$\text{agent } 3 \succ \text{ agent } 2$$
$$\text{agent } 3 \succ \text{ agent } 4$$

Now, in $\mathcal{L}_{13}$ defined in the procedure above, the allocation $(A = 2, B = 1)$ has strictly higher value than any other allocation in which the agent 3 is present. So,

$$\text{agent } 1 \succ \text{ agent } 3.$$

Thus,

$$\text{agent } 1 \succ \text{ agent } 3 \succ \text{ agent } 2 \text{ and}$$
$$\text{agent } 1 \succ \text{ agent } 3 \succ \text{ agent } 4$$

In $\mathcal{L}_{24}$, the allocation $(A = 2, B = 1)$ has strictly higher value than any other allocation in which the agent 4 is present. Thus, the ranking of the agents is,

$$\text{agent } 1 \succ \text{ agent } 3 \succ \text{ agent } 2 \succ \text{ agent } 4$$

It can be seen that the ranking defined above, satisfies the following properties.

1. $\succcurlyeq$ defines a total order on the set of bids.

2. $\succcurlyeq$ is independent of the order of the objects.

3. If two bids are the same, then they are equivalent in this order.

4. By increasing a bid, no agent will decrease his rank.

If agent $i \succcurlyeq$ agent $j$, we will also say $v_i \succcurlyeq v_j$.

## Appendix B. Some Proofs

### B.1 Proof of Lemma 1

- Suppose an optimal allocation contains an agent whose bid for his winning object, say $j$, is not in the top $p$ bids for the $j^{th}$ object. There are other $(p-1)$ winners in an optimal allocation. So, there exists at least one agent whose bid is in the top $p$ bids for the $j^{th}$ object and does not win any object. Thus, allocating him the $j^{th}$ object, we have an allocation which has higher valuation than the declared optimal allocation.

- Suppose an agent $i$ who receives an object in an optimal allocation is removed from the system. The agent will have at most one bid in the top $p$ bids for each object. So, agents now having bids in the top $p$ bids, will be at the $p^{th}$ position. It can be seen that there will be at most one agent in an optimal allocation who is on the $p^{th}$ position for the object he wins. If there is more than one agent in an optimal allocation on the $p^{th}$ position for the object they win, then we can improve on this allocation. Hence, after removing $i$, there will be at most one more agent who will be a part of a new optimal allocation.

$\square$

### B.2 Proof of Lemma 2

The argument is as follows.

1. Sort the bids of the agents for each object.

2. The optimal allocation consists of agents having bids in the $p$ highest bids for each of the objects (Lemma 1).

3. For computing the Clarke payment of the agent $i$, we remove the agent and determine an optimal allocation. And, using his bid, the valuation of optimal allocation with him and without him will determine his payment. This is done for each agent $i$. As per Lemma 1, if any agent from an optimal allocation is removed from the system, there exists a new optimal allocation which consists of at least $(p-1)$ agents who received the objects in the original optimal allocation.

4. There will be $p$ agents receiving the objects and determining their payments will involve removing one of them at a time, there will be at most $p$ more agents who will influence the payment. Thus, there are at most $2p$ agents involved in determining the Clarke payment.

□